\documentclass[fleqn]{annalen}
\usepackage{graphicx}
\usepackage[ansinew]{inputenc} 
\begin{document}

\newcommand{\volume}{9}              
\newcommand{\xyear}{2000}            
\newcommand{\issue}{1}               
\newcommand{\recdate}{dd.mm.yyyy}    
\newcommand{\revdate}{dd.mm.yyyy}    
\newcommand{\revnum}{0}              
\newcommand{\accdate}{dd.mm.yyyy}    
\newcommand{\coeditor}{ue}           
\newcommand{\firstpage}{1}           
\newcommand{\lastpage}{5}            
\setcounter{page}{\firstpage}        

\newcommand{\keywords}{Superluminal signal velocity, special relativity} 
\newcommand{\PACS}{03.30.+p,42.25Bs,73.40.GK,42.50.-p}
\newcommand{\shorttitle}{G. Nimtz, A. Haibel, Basics of Superluminal Signals} 

\title{Basics of Superluminal Signals} 

\author{G.~Nimtz and A. Haibel}
\newcommand{\address}{II.~Physikalisches Institut der Universit\"at zu 
K\"oln, Z\"ulpicher Str. 77, 50937 K\"oln, Germany }
\newcommand{\email}{\tt g.nimtz@uni-koeln.de} 

\maketitle

\begin{abstract}
The paper elucidates the physical basis of experimental results on superluminal signal velocity. 
It will be made plausible that superluminal signals do not violate the principle of causality but they 
can shorten the luminal vacuum time span between cause and effect. 
This amazing behaviour is based on the 
property that any physical signal has a finite duration. 
\end{abstract}

\section{Introduction}

One of the first investigations on wave propagation was presented by Lord Rayleigh at the end of the 
19th century. He considered that the group velocity corresponds to the velocity of energy or signal. 
This however raised difficulties with the theory of relativity in dispersive media. The problem was 
resolved by Sommerfeld and Brillouin in the case of waves with real  wave numbers, however, it was
not tackled for signals composed of evanescent modes only~\cite{Brillouin,Nimtz1}. 
For instance in  a textbook on {\it Relativity, Groups, Particles}
published recently a  chapter is devoted to signal velocities {\it faster than light}
~\cite{Sexl}. The authors conclude that superluminal group velocities may have been 
measured, however, they state {\it the initial packet gets completely deformed and unsuitable for perfect 
signal transmission during the course of propagation due to the vastly differing phase velocities of its
various frequency components.} 
The authors claim that a signal velocity never ever can exceed the vacuum velocity of light.
Allegedly this has been proven by Einstein studying the light  propagation in vacuum a hundred 
years ago in his famous paper~\cite{Einstein}.

\begin{figure}[htbp]
\hspace*{2cm}   (a)\\
\begin{center}
      \includegraphics[width=0.7\textwidth]{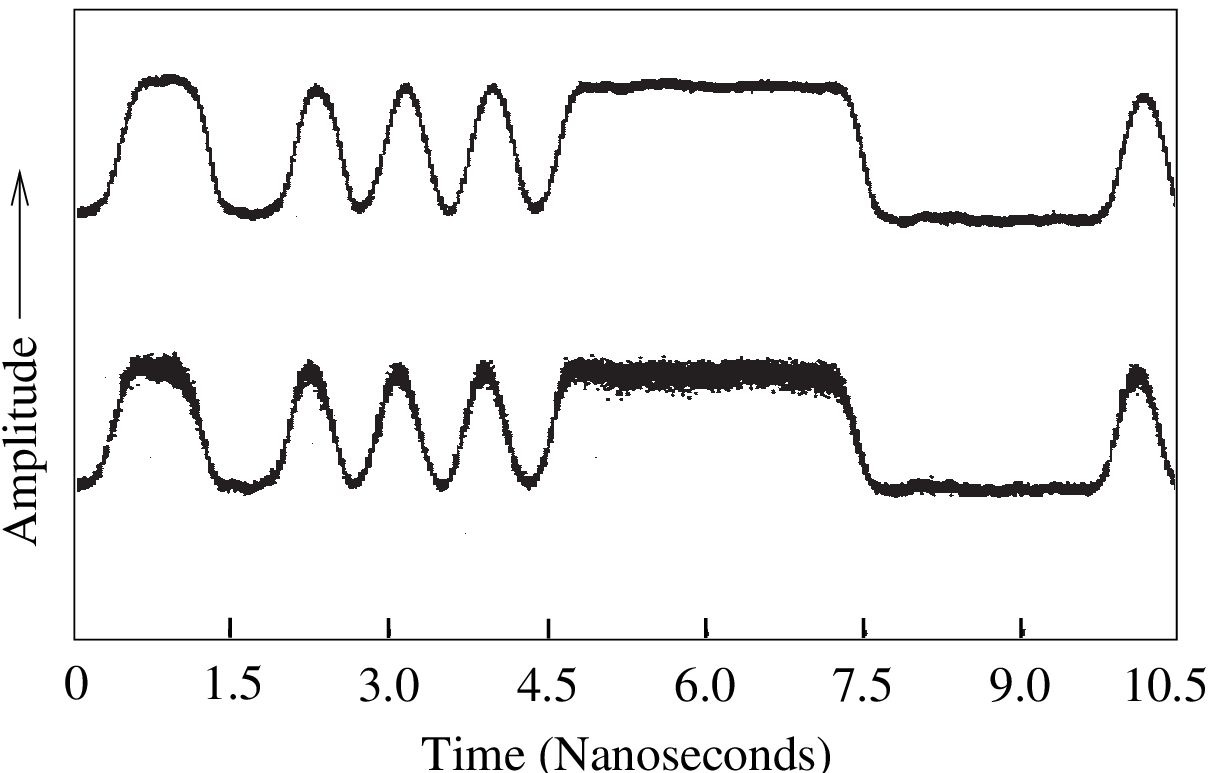}
\end{center}
\hspace*{2cm}   (b)\\
\begin{center}
      \includegraphics[angle=270,width=0.7\textwidth]{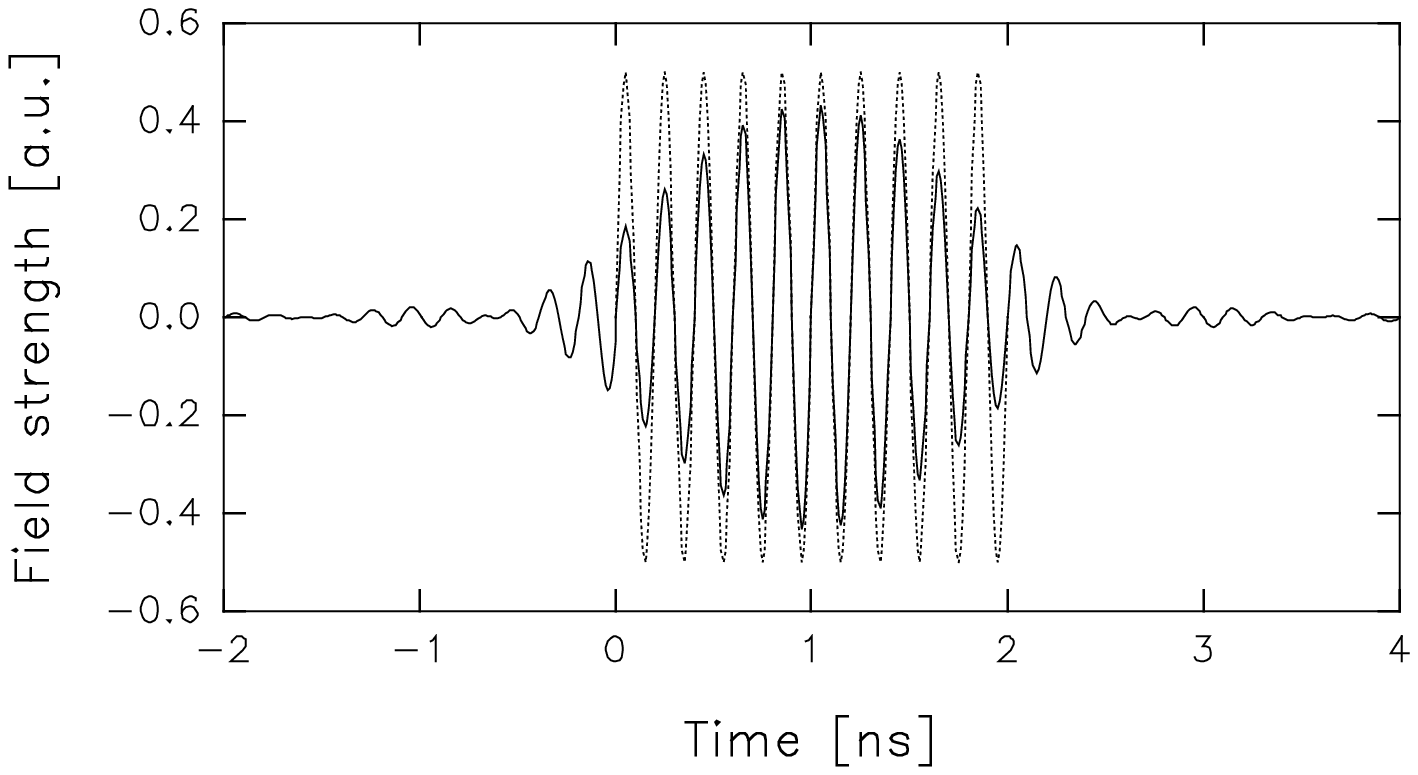}
\end{center}
      \caption{Signals: (a) Measured fiber optical signal intensity in arbitrary units (a.u.). 
                       The half width corresponds to the number of bits: 1,1,0,0,1,0,1,0,1.....
                       The lower signal was recorded after a distance of 9 000 km. The infrared 
                       carrier frequency of the signal is $2 \cdot 10^{14}$~Hz (wavelength 1.5$\mu$m). 
                       (b) Calculated sine wave signals; a mathematical non--frequency- (dotted line) 
                       and a physical frequency-band limited (solid line)~\cite{Brillouin}. As a result 
                       of the mathematical Fourier transform the frequency band limited signal has
                       already components at negative times, i.e. before it is switched on. Such a 
                       noncausal behaviour, however,  is not observed: Contrary to a mathematical signal 
                       a physical signal begins and ends gradually and causal as displayed on top of the 
                       figure. \label{bit}}
\end{figure}

In the case of tunnelling the above cited statement {\it vastly different phase velocity.....} 
is incorrect since an evanescent (tunnelling)
mode has a purely imaginary  wave number and does not experience a 
phase change~\cite{Sommerfeld,Feynman,Merzbacher,Nimtz2}. Therefore  the phase does not 
depend on the field spreading  inside a barrier. 
This outstanding property of the tunnelling process can indeed result in a superluminal signal velocity 
as will be reported in this article~\cite{Nimtz2,Nimtz3}.
Other books and papers even deny absolutely the possibility of either superluminal energy or 
signal velocities, see for instance references~\cite{Chiao,Sexl2,Goenner,Recami}. Some  authors 
claim that {\it frequency-band limitation is of course a technical and not a fundamental 
limitation}~\cite{Buettiker}. Yes certainly,  there is a finite probability for higher frequency 
components, even if their energy $\hbar \omega$ exceeds tremendously 
the signal's total energy~\cite{Recami}. However, these high frequency components having 
an extremely low probability and short lifetime are 
not traceable at all and their  statistical nature make fluctuations not suitable for signalling. 
Technical signals have usually a frequency band width much less than 1 $\%$ of the carrier frequency 
thus dispersion effects as a result of an interaction between the electromagnetic wave and any potential 
don't necessarily cause significant signal reshaping. 
For the extensive discussion and the various definitions of a signal  see e.g. 
Refs.~\cite{Nimtz3,Signal,HP,Smith}.
For example a digital signal as displayed in Fig.~\ref{bit}a is defined by the envelope which 
contains the total information. The amplitude don't carry information only the half width represents 
the  number of digits. Here the carrier frequency is 2$\times$10$^{14}$~Hz and the relative frequency
band modulation is 10$^{-4}$. For lang distance signal transmission signals were modulated on a high 
frequency carrier.

\begin{figure}[htb]\includegraphics[clip=,width=0.54\textwidth]{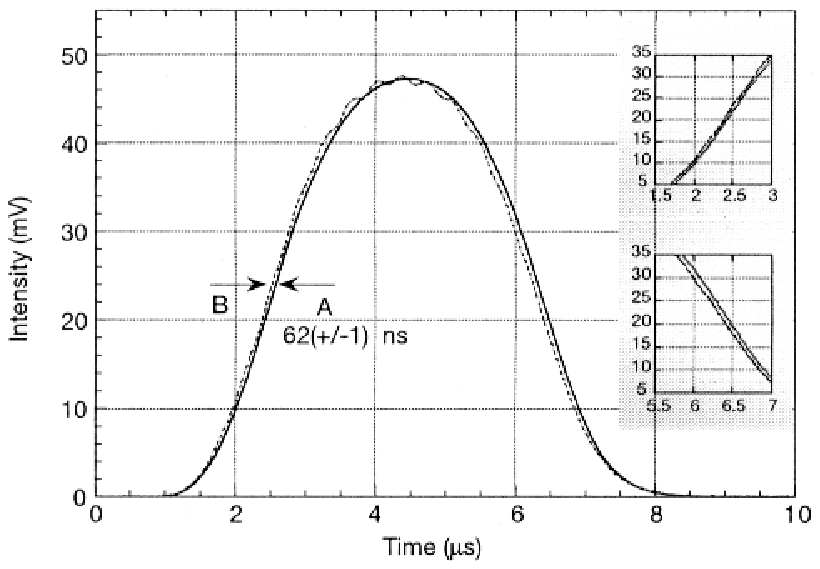}    
\hspace*{0.5cm}\includegraphics[width=0.45\textwidth]{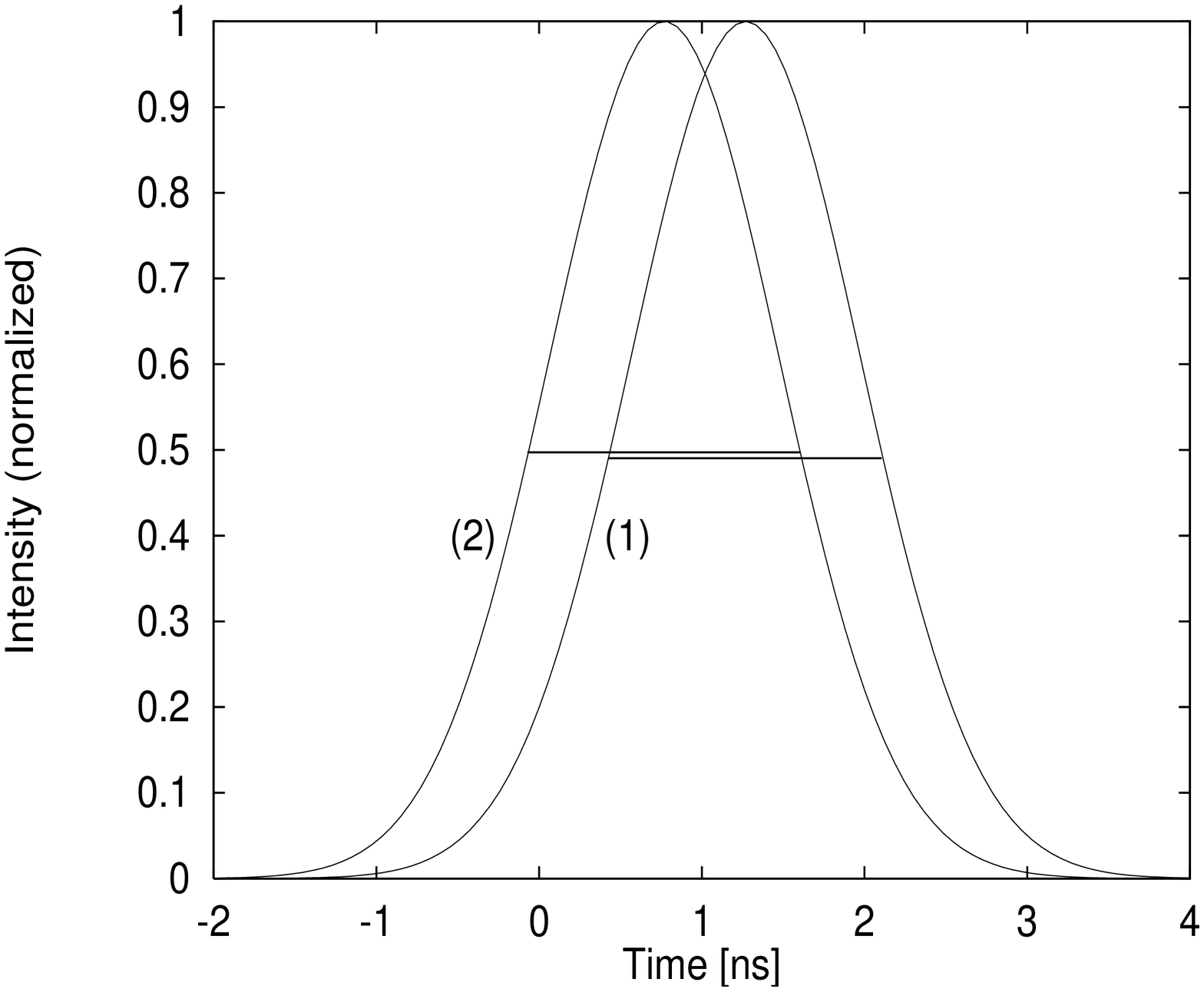} 
\caption{Display of superluminal gain--assisted optical (left~\cite{Wang})  
and  tunnelled microwave pulses (right~\cite{Nimtz4}).  The pulses (= signals) are normalized  
(a signal does not depend on its magnitude) and compared with 
the air born $A$ or the wave--guided  ones (1). 
 The measured time 
shifts of 62~ns and of  0.4~ns result in signal velocities of $-c/310$  and  of $4.7~c$, respectively.  
\label{pulse}}
\end{figure}

Figure~\ref{pulse}  presents two measured superluminal signals. They are pulse-like  
shaped as we are used to from digital communication systems. One pulse 
has travelled at a negative group velocity of $v_g$ = -c/310 the other one has tunnelled a
photonic barrier at a speed of $v_g$ = $4.7~c$, where $c$   
is the velocity of light in vacuum. The reader should inspect the two superluminal signals whether 
they are {\it completely deformed} as claimed in Ref.~\cite{Sexl}. A deformation of the signals is not 
seen and both  will be detected and interpreted properly as a well defined signal by the detector in 
question. In digital signal transmission the half width of the pulse represents the information 
(see Fig.~\ref{pulse} right and Fig.~\ref{bit}(a)). The superluminal time shift of the signal's envelope 
measured in both experiments represents the information that a superluminal signal including its 
energy have been measured. The signals shown in Fig.~\ref{pulse} have a frequency--band width 
of 120 kHz at a carrier frequency of  350 THz and 300 MHz at a carrier frequency of 10 GHz,  
respectively.

\begin{figure}
    \includegraphics[width=\linewidth]{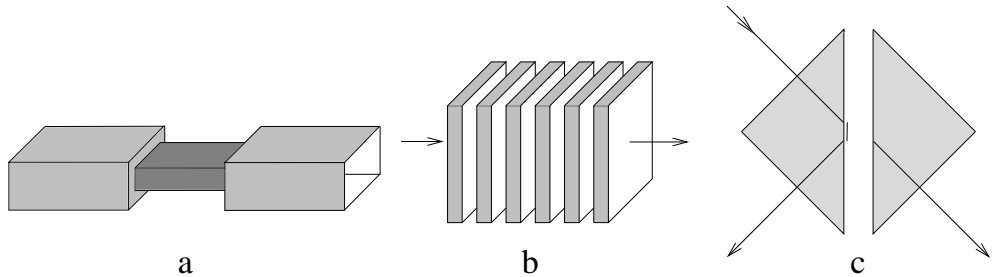}
\caption{Sketch of the three prominent photonic barriers. a) illustrates an undersized waveguide 
(the central part of the wave guide has a cross-section being smaller than half the wavelength 
in both directions perpendicular to propagation) , b) a photonic lattice (periodic dielectric hetero 
 structure), and c) the frustrated total internal reflection of a double prism, where total reflection 
takes place at the boundary from a denser to a rarer dielectric medium.}
\label{Barrier}
\end{figure}

Obviously superluminal signal velocities do exist. As an example of superluminal signal transmission 
the tunnelling process is discussed here. Electromagnetic waves with a purely  imaginary wave 
number are called evanescent modes. They play an important role in microwave technology and in 
classical optics. As a result of their mathematical analogy with wave mechanical tunnelling the 
instantaneous process of the spreading of evanescent fields is called photonic 
tunnelling~\cite{Sommerfeld,Feynman}.  Three prominent photonic barriers for studying the 
tunnelling process are presented in Fig.~\ref{Barrier}.

The article begins with a presentation of the special properties of a physical signal. Amazingly 
enough in the following section it is shown that superluminal signal velocities can shorten the 
luminal time span between cause and effect but do not violate the principle of causality. 

\section{Signals} 

A signal is thought to cause a  corresponding effect. Basically phonons, photons, and electrons 
are exploited to mediate interaction or to transmit bits, words or any desired signal to induce 
requested effects. The front or the discontinuous beginning of a signal  is only well defined 
in the mathematical case of an infinite frequency spectrum. Any physical transmitter produces 
signals of finite spectra only. Therefore a front has no physical meaning and the narrow--band 
and the wide--band envelope (defined and discussed in Ref.~\cite{Brillouin,Papoulis}) 
is the appropriate signal description. An example is displayed in Fig.~\ref{bit}a. In the case of an 
infinite signal spectrum the Planck quantum $\hbar \omega$ of the minimum energy of a field 
would necessarily result in an infinite signal energy as mentioned in the introduction. 

A signal may be a single photon or electron with a distinct energy, more general
a signal is characterized by a carrier frequency and its modulation and it is bound to be 
independent of magnitude otherwise we could not listen to the radio broadcasting or mobile 
phone during driving a car. There are frequency (FM) and amplitude (AM) modulated signals 
or a combination of both. A signal is not described by an analytical function, otherwise the 
complete information would be contained in the forward tail of a signal. In the latter case a 
detector could recognize from the forward tail the rest of the shape of the signal~\cite{Low}. 
This becomes  obvious for instance in digital optoelectronic communication systems, where 
measuring the signal half width gives the number of digits. Such a digital signal sent via glass 
fiber in modern communication systems is presented in Fig.~\ref{bit}(a). 

In the same figure part \ref{bit}(b) is a calculated sine wave signal of 10 oscillations plotted.  
The dotted one has an unlimited frequency--band width, whereas the solid one has a finite 
frequency--band. The carrier frequency is 5~GHz and the width 500~MHz. As a result of 
the frequency--band limitation there are noncausal components. The calculated result follows 
from the mathematical Fourier transform of a frequency band limitation.  The noncausal 
components are not measurable. Any  physical signal is beginning gradually at the time $t = 0$  
as seen in Figs.~\ref{pulse}, \ref{bit}(a)~\cite{Brillouin,Nimtz3,Papoulis}. A discontinuously  
signal beginning would make necessary  the existence of infinite frequency components.

In the following we repeat the definitions of velocities which are found presented extensively 
elsewhere, e.g. in Refs.~\cite{Brillouin,Feynman,Merzbacher,Papoulis,Stratton}:

\begin{eqnarray}
v_{\varphi}  &=& \omega /k \\
v_g &=& d\omega /dk\\
v_s & \equiv & v_g,
\label{vsvg}
\end{eqnarray}
 where $\omega$ and k are the angular frequency and the wave number, respectively. $v_\varphi$ 
is the phase, $v_g$ is the group,  and $v_s$ is the signal velocity.  
Relation (3) of the signal velocity $v_s$ is valid in vacuum or in other anomalous dispersion--free 
media.

\begin{figure}[htb]
\center{\includegraphics[width=0.6\textwidth]{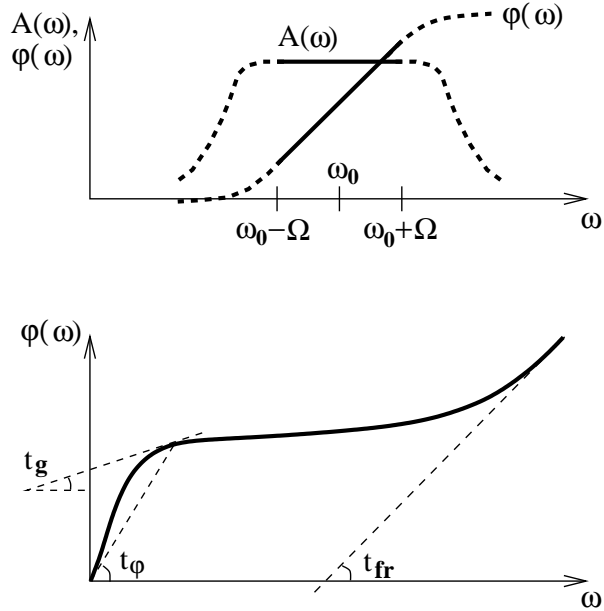}    }
        \caption{A sketch of a signal's frequency-bandwidth $A(\omega)$ (central frequency 
                          $\omega_0 \pm \Omega$),  of phase angle $\varphi(\omega)$, and of 
                         various delay times as defined e.g. in Ref.~\cite{Papoulis}. $\Omega$ is 
                         the frequency-band width $\omega_0 \pm \Omega$ around the center 
                         frequency $\omega_0$. \label{Pap}}
\end{figure}

For the problem of signal propagation the following terms are used to describe the delay of the 
various parts of a signal envelope. The delay times have been analyzed for instance in the text book 
on Fourier Transform by Papoulis~\cite{Papoulis}:  
\begin{eqnarray}
t_{\varphi}(\omega) &=& \varphi(\omega) / \omega 
\label{delaytime1} \\
t_g(\omega) &=& d\varphi(\omega) /d \omega 
\label{delaytime2} \\
t_{fr}(\omega) &=& \lim_{\omega \rightarrow \infty} \varphi(\omega) / \omega
\label{delaytime3}
\end{eqnarray}
where $t_{\varphi}$ is the phase time delay, $t_g$ is the group time delay,  and $t_{fr}$ is the 
front time delay,  see Fig.~\ref{Pap}~\cite{Papoulis}. $\varphi = k z$ is the phase in a medium 
at length $z$. The frequency--band $A(\omega)$, the phase angle $\varphi$, and the delay times 
are illustrated in Fig.~\ref{Pap}. If  $\varphi$ does not tend to a straight line as $\omega$ tends to 
infinity, then the term signal front delay has no meaning~\cite{Brillouin,Papoulis}. This behaviour 
takes place in the case of frequency--band limited signals in any medium and  in the case of tunnelling, 
where the wave number is imaginary and thus the angle $\varphi$ does not depend on length $z$. 
If the signal is narrow-band limited, there is no distortion of the envelope. Therefore the delay of 
the signal envelope is assumed to equal the delay of the center of gravity~\cite{Papoulis}. 

The classical {\it luminal forerunners}~\cite{Brillouin,Nimtz1,Stratton} don't exist in the case of a 
tunnelling signal containing evanescent modes only. Luminal forerunners exist only in dispersive 
media with a finite real refractive index.

\section{Superluminal Signals}

Since Helmholtz and Schr\"o\-dinger equations are mathematically identical, it is evident that 
the three kinds of barriers displayed in Fig.~\ref{Barrier} can be used to model the 
one--dimensional process of wave mechanical tunnelling~\cite{Sommerfeld,Feynman}.

Some superluminal data obtained with the historical double--prism experiment are presented 
in the following~\cite{Sommerfeld,Feynman,Haibel,Nimtz5}. This smart experiment elucidates 
in a brilliant way the time  behaviour of the tunnelling process~\cite{Stahlhofen}. 

\begin{figure}[htb]
          \center{ \includegraphics[width=0.35\textwidth]{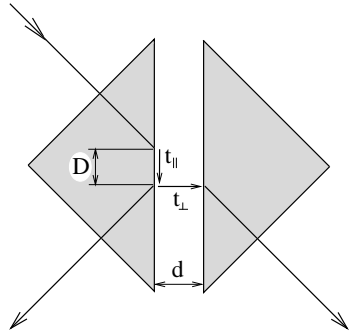}
          \caption{The tunnelling time of the double-prism experiment consists of two 
                            components. $t_{\|}$ for the Goos--Hänchen shift $D$ parallel to 
                            the prism's surface and $t_{\bot}$ for crossing the gap in the direction 
                           perpendicular to the two surfaces of the gap. \label{doubleprism} } }  
\end{figure}

Figure~\ref{doubleprism} indicates that for an angle of incidence greater than the angle of total 
reflection the barrier transmission time of the double--prism, or what we call here the tunnelling 
time can be split into two components 

\begin{eqnarray}
t_{\rm tunnel}=t_{\|} + t_{\bot},
\end{eqnarray}

\noindent
one along the surface due to the Goos-Hänchen shift $D$, and another part perpendicular to the 
surface~\cite{Haibel,Stahlhofen}. The measured tunnelling time $t_{tunnel}$ represents the 
group time delay which results in the group or signal velocity (see Eq.~(\ref{vsvg})).

The first component of $t_{tunnel}$ is related to a non-evanescent wave characterized 
by the real  wavenumber

\begin{eqnarray}
 k_{\|}\,:=\,k_0\,n_1\,\sin \theta_i
\end{eqnarray}

\noindent
while the second one 

\begin{eqnarray}
k_{\bot}\,:=\,i\,k_0\,\sqrt{n_1^2\,\sin^2\,\theta_i\,-\,1}
\end{eqnarray}

\noindent
is related to the evanescent mode traversing the gap between the two prisms. 
$k_0 = 2\pi/\lambda_0$, where $\lambda_0$ is the corresponding vacuum wavelength, 
and $n_1$ the refractive index of both prisms. Details of the experiment are given 
in~\cite{Haibel}.

In the symmetrical design of the experiment displayed in Fig.~\ref{doubleprism} the reflected 
and the tunnelled signal leave the first and the second prisms at the same time. This result 
represents an experimental proof that for the tunnelling time component  $t_{\bot}$ holds  
$t_{\bot}$ = 0. It is in agreement with the observations on other photonic tunnelling structures: 
The finite measured tunnelling time accrues at the entrance boundary and no time is spent inside 
a barrier.

A couple of months after the discovery of  superluminal tunnelling of microwave signals~\cite{Enders} 
a study on superluminal group velocity and transmission of single optical photons tunnelling a photonic 
barrier was published~\cite{Steinberg}. Yes certainly, as the authors claimed they did not measure a 
signal velocity as the photons were emitted in a spontaneous process. However, the group velocity of the 
investigated black box (i.e. the tunnelling barrier) has  been determined and the data are also valid  for 
the transmission of signals. Sending signals containing millions of optical photons analogous to  the 
microwave experiment, the black box would result in the same superluminal group velocity as in the 
single photon experiment. In fact the same experimental set-up and procedure have been proven with 
a sample of bulk glass instead of the tunnelling barrier this black box analogy to be 
correct~\cite{Steinberg2}. Actually in this case the single photon experiment yielded the 
{\it sub-}luminal group velocity known from bulk glass as measured in standard spectroscopy. 
 In both in the microwave and in the single photon experiments the group velocity has been measured 
with a detector located in free space far away from the investigated black box. In such asymptotic 
measurements the relation (3) holds, i.e. the group velocity equals the signal velocity.

\section{Does Superluminal Signals Violate\\
The Principle Of Causality?}

\begin{figure}[htb]
\center{
\includegraphics[width=0.6\textwidth]{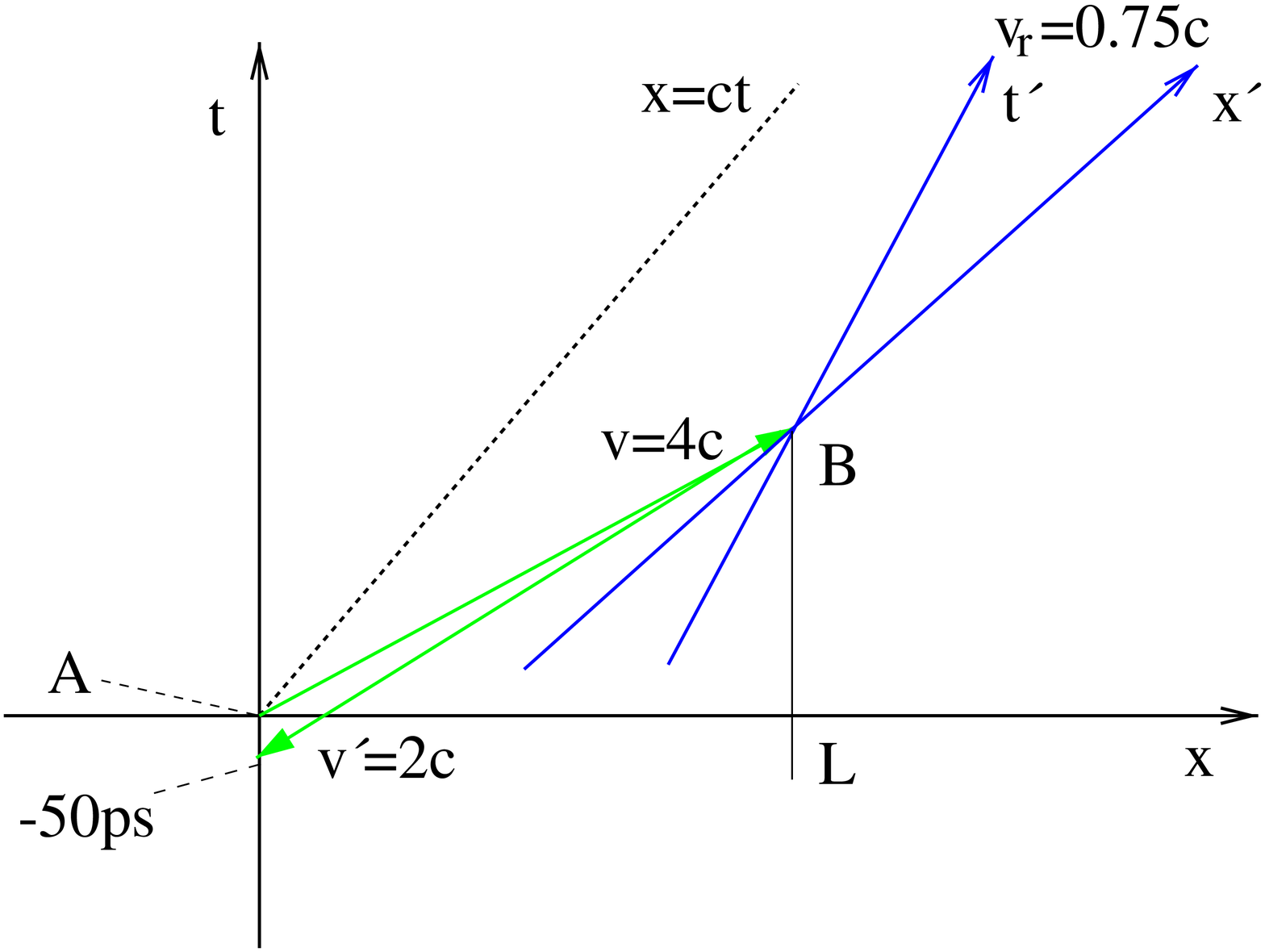}    } 
\caption{Coordinates of two observers {\bf A} $(0,0)$ and {\bf B} with $O(x,t)$ and $O'(x',t')$ moving
                with a relative velocity of $0.75~c$. The distance $L$ between {\bf A} and {\bf B}   is 0.1~m.      
                {\bf A} has available  a signal velocity $v_s = 4~c$  and {\bf B}  $v_s'= 2~c$. The numbers in the 
                example are chosen according to~\cite{Mittelstaedt}.  The signal returns -50~ps in the past 
                 in {\bf A}.
\label{koord}}
\end{figure}

Photonic tunnelling experiments have revealed superluminal signal 
velocities~\cite{Nimtz4,Haibel,Enders} as well as superluminal group and energy 
velocities~\cite{Steinberg}  of single photons. According to textbooks a superluminal signal 
velocity violates Einstein causality implying that cause and effect can be changed and time 
machines known from science fiction  could be constructed. This interpretation, however, 
assumes a signal to be a point in the time dimension neglecting its finite duration. Can a signal 
travelling faster than light really violate the principle of causality stating that cause precedes 
effect~\cite{Mittelstaedt}?  The latter question has been widely assumed as a matter of fact and 
it has allegedly been shown that a signal velocity faster than light allows to change the past. The 
line of arguments how to manipulate the past in this case is illustrated in  
Fig.~\ref{koord}~\cite{Sexl,Chiao,Sexl2,Goenner,Recami}. There are two frames of reference 
displayed. In the first one at the time $t = 0$  lottery numbers are presented, whereas at 
$t = -10~$ps the counters were closed. Mary ($A$) sends the lottery  numbers to her friend 
Susan ($B$) with a signal velocity of 4 $c$. Susan, moving in the second inertial system at a 
relative speed of 0.75~$c$, sends the numbers back at a speed of  2 $c$, which arrives in the first 
system at $t = -50~$ps, thus in time to deliver the correct lottery numbers before the counters 
close at  $t = -10~$ps.

The time shift of a point on the time coordinate of the reference system $A$ into 
the past is given by the relation~\cite{Sexl2,Mittelstaedt}:

\begin{eqnarray}
t_A = - \frac{L}{c} \cdot \frac{(v_r - c^2/v_s - c^2/v_s' + c^2 v_r/ v_s v_s')}{(c - c v_r/ v_s')}
\end{eqnarray}

where $L$ is the transmission length of the signal, $v_r$ is the 
velocity between the two inertial systems $A$ and $B$. 
The condition 
for the change of chronological order is $t_A \le 
0$, the time shift between the systems $A$ and 
$B$. 

\begin{figure}[htb]
\center{\includegraphics[width=0.6\textwidth]{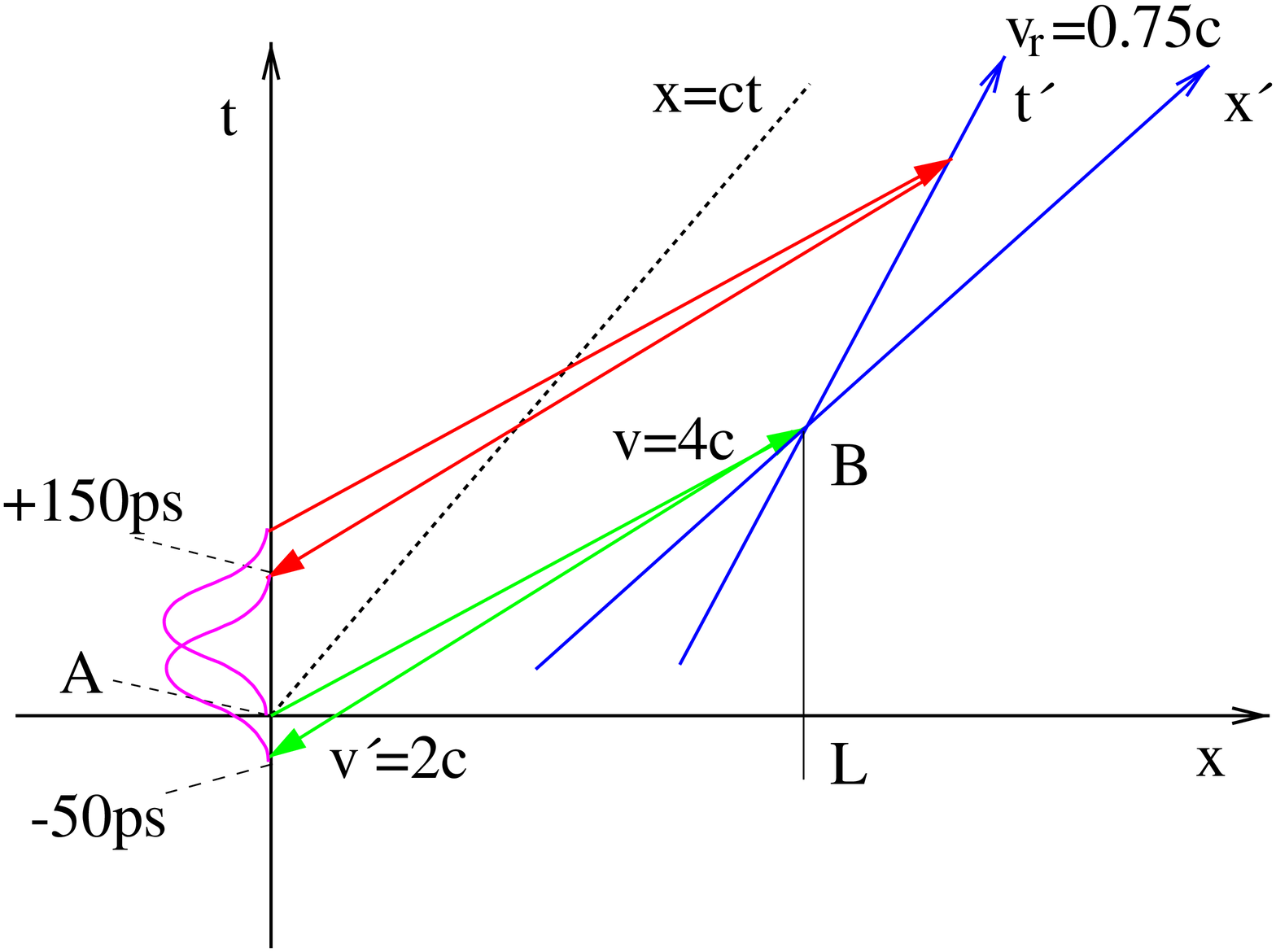}    } 
\caption{In contrast to Fig.~\ref{koord} the pulse has now a finite duration of $200$ ps.
                                    This data is used  for a clear demonstration of the effect. In both experiments,   
                the signal length is long  compared with the measured negative time shift in 
                  spite of the superluminal signal velocity. A signal envelope ends generally in the future. 
\label{koord1}}
\end{figure}

In the experiments displayed in Fig.~\ref{pulse} 
the signals travelled at a superluminal velocity 
with $-c/310$~\cite{Wang}
and with 
$4.7~c$~\cite{Nimtz4} 
, respectively. Nevertheless, the principle of causality has not been violated 
in both experiments. In the example with the lottery data the signal was assumed to be a point on the 
time coordinate. 
However, a signal has a finite duration as the pulses  sketched along the time coordinate in 
Fig.~\ref{koord1}.  

\begin{figure}[htb]
\center{\includegraphics[width=0.5\textwidth]{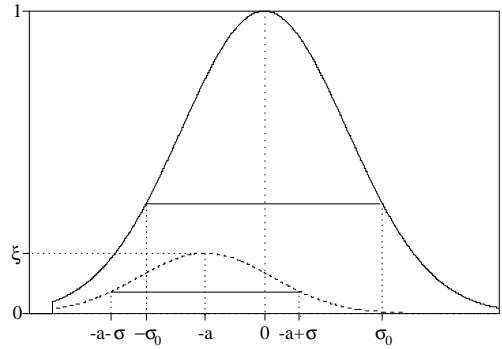}    } 
\caption{
Comparison of normalized intensity vs time of an air born signal (solid line) and a simulated tunnelled 
signal (dotted line) moving from right to left. The signals have a discontinuous beginning 
and the frequency band is unlimited. The tunnelled pulse is attenuated and reshaped. Moreover, 
although its maximum has 
travelled at superluminal speed, both fronts have traversed the same distance with the light velocity 
in vacuum.  
Here $\xi$ is the maximum of the tunnelled pulse, a is the shift of the maximum, $\sigma$ is the 
variance of the tunnelling signal, and $\sigma_0$ is the variance of the air born pulse. 
It is obvious that the latter is longer than the variance of the tunnelled pulse.
\label{Emig}}
\end{figure}

The two signals presented in Fig.~\ref{pulse} have lengths of $7~\mu$s and 4~ns.  
Any signal like a bit, a word  or a sentence has a finite extension on the time coordinate. 
Assuming in our example a signal's finite duration 
of  200~ps the complete information is obtained with superluminal velocity but only at positive times 
under the assumptions  illustrated in Fig.~\ref{koord1}. 
The same holds a fortiori for the two presented experiments. The finite duration of a signal is the reason  
that a superluminal velocity does not violate the principle of causality. A shorter  signal with the same 
information (such a transform is done frequently 
in modern transmission technology) has an equivalently broader frequency band. 
In this case as a result of  the dispersion relations of tunnelling barriers or of any interaction 
a pulse reshaping would take place.  Such a pulse reshaping is demonstrated in Fig.~\ref{Emig}, details 
are presented in Ref.~ \cite{Emig}. 
 
The pulse reshaping is a result of a discontinuous leading front and an unlimited frequency--band. 
Both shape and half width are changed during the tunnelling process.

\section{Summing up}

Evanescent modes show some amazing properties to which we 
are not used to from classical physics.  Apparently evanescent modes are nonlocal fields and 
represented by virtuell photons~\cite{Cargnilia}. Evanescent modes are easily traceable through 
barriers some 100 wavelengths thick and they do not spend time in the barrier. The latter is an 
experimental result due to the fact that the transmission time is independent of 
barrier length (Hartman effect \cite{Hartman,Enders2,Spielmann}). Another proof of this behaviour 
is observed in the case of symmetrical  Frustrated Total Internal Reflection (FTIR), where the 
reflected and the transmitted signal have the same delay time, i.e. the time spent inside the barrier is 
zero. The measured finite transmission time comes into existence at the entrance boundary of the 
photonic barriers. 

In conclusion the principle of causality has not been violated  by superluminal signals 
as a result of the finite signal length and the corresponding frequency-band width.
But amazingly enough the time span between cause and effect is reduced by a superluminal 
signal velocity compared with luminal cause to effect propagation.

\section{Acknowledgment}

The author acknowledge gratefully the very helpful discussions with Cl. Laemmerzahl, P. Mittelstaedt, 
A. Stahlhofen, and R.-M. Vetter.

\end{document}